\documentclass[]{llncs}

\usepackage[utf8]{inputenc}
\usepackage{graphicx}
\usepackage{tabularx}
\usepackage{booktabs}
\usepackage{makecell}
\usepackage{todonotes}
\usepackage{booktabs}
\usepackage{subcaption}
\usepackage{tcolorbox}
\usepackage{wrapfig}
\usepackage{xcolor}
\usepackage{wasysym}
\usepackage[all]{nowidow}
\usepackage{url}
\usepackage{fancyvrb}

\def\rot{\rotatebox}

\colorlet{punct}{red!60!black}
\definecolor{background}{HTML}{EEEEEE}
\definecolor{delim}{RGB}{20,105,176}
\colorlet{numb}{magenta!60!black}

\usepackage{listings}
\lstdefinelanguage{json}{
    basicstyle=\normalfont\ttfamily,
    numberstyle=\scriptsize,
    stepnumber=1,
    numbersep=8pt,
    showstringspaces=false,
    breaklines=true,
    frame=lines,
    backgroundcolor=\color{background},
    literate=
     *{0}{{{\color{numb}0}}}{1}
      {1}{{{\color{numb}1}}}{1}
      {2}{{{\color{numb}2}}}{1}
      {3}{{{\color{numb}3}}}{1}
      {4}{{{\color{numb}4}}}{1}
      {5}{{{\color{numb}5}}}{1}
      {6}{{{\color{numb}6}}}{1}
      {7}{{{\color{numb}7}}}{1}
      {8}{{{\color{numb}8}}}{1}
      {9}{{{\color{numb}9}}}{1}
      {:}{{{\color{punct}{:}}}}{1}
      {,}{{{\color{punct}{,}}}}{1}
      {\{}{{{\color{delim}{\{}}}}{1}
      {\}}{{{\color{delim}{\}}}}}{1}
      {[}{{{\color{delim}{[}}}}{1}
      {]}{{{\color{delim}{]}}}}{1},
}

\graphicspath{ {figures/} }

\DeclareCaptionLabelFormat{andtable}{#1~#2  \&  \tablename~\thetable}

\begin{document}



\title{Overview of LiLAS 2021 -- Living Labs for Academic Search}

\author{Philipp Schaer\inst{1} \and 
        Timo Breuer\inst{1} \and 
        Leyla Jael Castro\inst{2} \and 
        Benjamin Wolff\inst{2} \and \\
        Johann Schaible\inst{3} \and
        Narges Tavakolpoursaleh\inst{3}}
\authorrunning{Schaer et al.}

\institute{TH Köln -- University of Applied Sciences, Cologne, Germany\\
\email{firstname.lastname@th-koeln.de}
\and
ZB~MED -- Information Centre for Life Sciences, Cologne, Germany\\
\email{{ljgarcia,wolff}@zbmed.de}
\and
GESIS -- Leibniz Institute for the Social Sciences, Cologne, Germany\\
\email{firstname.lastname@gesis.org}
}

\maketitle

\begin{abstract}
The Living Labs for Academic Search (LiLAS) lab aims to strengthen the concept of user-centric living labs for academic search. The methodological gap between real-world and lab-based evaluation should be bridged by allowing lab participants to evaluate their retrieval approaches in two real-world academic search systems from life sciences and social sciences. This overview paper outlines the two academic search systems LIVIVO and GESIS Search, and their corresponding tasks within LiLAS, which are ad-hoc retrieval and dataset recommendation. The lab is based on a new evaluation infrastructure named STELLA that allows participants to submit results corresponding to their experimental systems in the form of pre-computed runs and Docker containers that can be integrated into production systems and generate experimental results in real-time. Both submission types are interleaved with the results provided by the productive systems allowing for a seamless presentation and evaluation. The evaluation of results and a meta-analysis of the different tasks and submission types complement this overview. 

\end{abstract}

%

\section{Introduction}

The Living Labs for Academic Search (LiLAS) lab aims to strengthen the concept of user-centric living labs for the domain of academic search. By allowing lab \emph{participants} to evaluate their retrieval approaches in two real-world academic search portals (called \emph{sites}) from life sciences and social sciences, the methodological gap between real-world and lab-based evaluations is effectively reduced.

\begin{figure}[t]
\centering
\includegraphics[width=\textwidth]{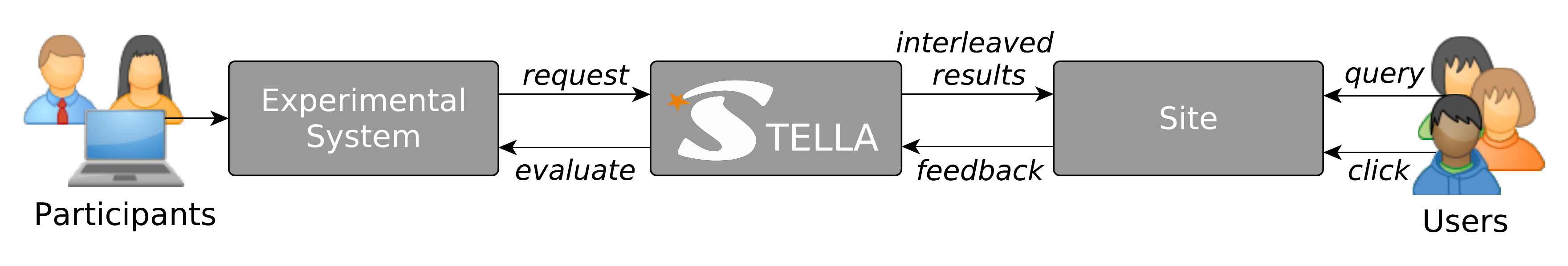}
\caption{Overview of the live evaluation pipeline.}
\label{fig:lilas-overview-simple}
\end{figure}

This gap is based on the different opportunities available to researchers in academia and industry. While industry-based research in the field of information retrieval (IR) has the opportunity to conduct experiments in-vivo -- thanks to the availability of large systems, with a wide range and correspondingly large user base -- these opportunities usually remain closed to academic research. In-vivo here describes the possibility to perform IR experiments integrated into real-world systems and to conduct experiments where the actual interaction with these systems takes place. It should be emphasized here that these are not classic user experiments in which the focus is on the individual interactions of users (e.g., to investigate questions of UI design), but rather aggregated usage data is collected in large quantities in order to generate reliable quantitative research results. 
The potential of living labs and real-world evaluation techniques has been shown in previous CLEF labs such as NewsREEL~\cite{lommatzsch2018newsreel} and  LL4IR~\cite{DBLP:conf/clef/SchuthBK15}, or TREC OpenSearch~\cite{balog2016overview}. In a similar vein, LiLAS is designed around the living lab evaluation concept and introduces different use cases in the broader field of academic search. Academic search solutions, which have to deal with the phenomena around the exponential growing rate \cite{de_solla_price_little_1963} of scientific information and knowledge, tend to fall behind the real-world requirements and demands. The vast amount of scientific information does not only include traditional journal publication, but also a constantly growing amount of pre-prints, research datasets, code, survey data, and many other research objects. This heterogeneity and mass of documents and datasets introduces new challenges to the disciplines of information retrieval, recommender systems, digital libraries, and related fields. Academic search is a conceptional umbrella to subsume all these different disciplines and is well-known through (mostly domain-specific) search systems and portals such as PubMed, arXiv.org, or dblp. While those three are examples of open-science-friendly systems as they allow re-use of metadata, usage data and/or access to fulltext data, other systems such as Google Scholar or ResearchGate. The later offer no access at all to their internal algorithms and data and are therefore representatives of a closed-science (and commercial) mindset. 

Progress in the field of academic search and its corresponding domains is usually evaluated by means of shared tasks that are based on the principles of Cranfield/TREC-style studies~\cite{Schaible2020}. 
Most recently the TREC-COVID 
evaluation campaign run by NIST attracted a high number of participants and showed the high impact of scientific retrieval tasks in the community. Within TREC-COVID a wide range of systems and retrieval approaches participated and generally showed the massive retrieval performance that recent BERT and other transformer-based machine learning approaches are capable of. However, classic vector-space retrieval was also highly successful 
and showed the limitations of the test collection-based evaluation approach of TREC-COVID and the general need for innovation in the field of academic search and IR. Meta-evaluation studies of system performances in TREC and CLEF showed a need for innovation in IR evaluation~\cite{armstrong_improvements_2009}. The field of academic 
search is no exception to this. The central concern of academic search is finding both relevant and high-quality documents. The question of what constitutes relevance in academic search is multilayered \cite{DBLP:conf/ecir/CarevicS14} and an ongoing research area.

In 2020 we held a first iteration of LiLAS as a so-called workshop lab. This year we provide participants exclusive access to real-world systems, their document base (in our case a very heterogeneous set of research articles and research data including, for instance, surveys), and the actual interactions including the query string and the corresponding click data (see overview on the setup in Figure \ref{fig:lilas-overview-simple}). To foster different experimental settings we compile a set of head queries and candidate documents to allow pre-computed submissions. Using the STELLA-infrastructure, we allow participants to easily integrate their approaches into the real-world systems using Docker containers and provide the possibility to compare different approaches at the same time.

This lab overview is structured as follows: In Sections \ref{sec:usecase-livivo} and \ref{sec:usecase-gesis} we introduce the two main use cases of LiLAS which are bond to the sites granting us access to their retrieval systems: LIVIVO and GESIS Search. In these two sections the systems, the provided datasets, and task are described. In Section \ref{sec:setup} we outline the evaluation setup and STELLA, our living lab evaluation framework, and the two submission types, namely pre-computed runs and Docker container submissions. Section \ref{sec:setup} also includes the description of the evaluation metrics used with in the lab and a short overview on the organizational structure of the lab. In Section \ref{sec:participation} we introduce the participating groups and approaches. We outline the results of the evaluation rounds in Section \ref{sec:results} and conclude in Section~\ref{sec:conclusion}.

\section{Ad-hoc Search in LIVIO}
\label{sec:usecase-livivo}

\subsection{LIVIVO Literature Search Portal}
LIVIVO\footnote{\url{https://www.livivo.de}} \cite{muller_livivo_2017} is a literature search portal developed and supported by ZB MED – Information Centre for Life Sciences.
ZB MED is a non-profit organization providing specialized literature in Life Sciences at a national (German) and international level and hosting one of the largest stock of life science literature in Europe. Since 2015, ZB MED supports users including librarians, students, general practitioners and researchers with LIVIVO, a comprehensive and interdisciplinary search portal for Life Sciences.

LIVIVO integrates various literature resources from medicine, health, environment, agriculture and nutrition, covering a variety of scholarly publication types (e.g., conferences, preprints, peer-review journals). LIVIVO corpus includes about 80 million documents from more than 50 data sources in multiple languages (e.g., English, German, French). To better support its users, LIVIVO offers an end-user interface in English and German, an automatically and semantically enhanced search capability, and a subject-based categorization covering the different areas it supports (e.g., environment, agriculture, nutrition, medicine). Precision of search queries is improved by using descriptors with semantic support; in particular, LIVIVO uses three multilingual vocabularies to this end (Medical Subject Headings MeSH,
UMTHES,
and AGROVOC.
In addition to its search capabilities, LIVIVO also integrates functionality supporting inter-library loans at a national level in Germany. 
Since 2020, LIVIVO also offers a specialized collection on COVID-19\footnote{\url{https://www.livivo.de//covid19}.
}

\subsection{LIVIVO Dataset}

\begin{figure}[t]
\input{sections/livivo-code}
\caption{Examples for head queries, documents, and candidate lists for the LIVIVO system.}
\label{listing:livivodata}
\end{figure}
%
%
%

For the LiLAS challenge, we prepared training and test datasets comprising head queries together with 100-document candidate list. In Figure \ref{listing:livivodata} we include an excerpt of the different elements included in the data. Data was formatted in JSON and presented as JSONL files to facilitate processing. Participating head queries were restricted to keywords-based search and keywords-based search plus AND, OR and NOT operators.

Head queries were assigned an identifier, namely \emph{qid}, a query string, \emph{qstr} and as an additional information the query frequency, \emph{freq}. For each head query, a candidate list was also provided. Candidate lists include the query identifier as well as corresponding string, together with a list of 100 document identifiers (i.e. the native identifier used in the LIVIVO database). 

In addition to head queries and candidate lists, we also provided a set of documents in LIVIVO corresponding to three of the major bibliographic scholarly databases so participants could create their own indexes.
The document set contains metadata for approx. 35 million documents and is provided as a JSONL file. To reduce complexity and keep the data manageable, we decided to provide only the 6 most important data fields (DBRECORDID, TITLE, AUTHOR, SOURCE, LANGUAGE, DATABASE). Additional metadata and fulltext is mostly available from the original database curators. 
The aformentioned databases correspond to Medline,
the National Library of Medicine's (NLM) bibliographic database for life sciences and biomedical information including about 20 million of abstracts; the NLM catalog,
providing access to bibliographic data for over 1.4 million journals, books and similar data; and the Agricultural Science and Technology Information (AGRIS) database,
a Food and Agriculture Organization of the United Nations initiative compiling information on agricultural research with 8.9 million structured bibliographical records on agricultural science and technology. 

%

\subsection{Task}

Finding the most relevant publications in relation to a head query remains a challenge in scholarly Information Retrieval systems. While most repositories or registries deal mostly with publications in English, LIVIVO, the production system used at LiLAS, supports multilingualism, adding an extra layer of complexity and presenting a challenge to participants.

The goal of this ad-hoc search task is supporting researchers to find the most relevant literature regarding a head query. Participants were asked to define and implement their ranking approach using as basis a multi-lingual candidate documents list. A good ranking should present users with the most relevant documents on top of the result set. An interesting aspect of this task is the multilingualism as multiple languages can be used to pose a query (e.g. English, German, French); however, regardless of the language used on the query, the retrieval can include documents in other languages as part of the result set.

\section{Research Data Recommendations in GESIS-Search}
\label{sec:usecase-gesis}

\subsection{GESIS Search Portal}\label{sec:gesissearch}
GESIS Search\footnote{\url{https://search.gesis.org/}} is a search portal for social science research data and open access publications developed and supported by 
GESIS - Leibniz Institute for the Social Sciences. 
GESIS is a member of the Leibniz Association with the purpose to promote social science research.
It provides essential and internationally relevant research-based services for the social sciences, and as the largest European infrastructure institute for the social sciences, GESIS offers advice, expertise and services to scientists at all stages of their research projects. 

GESIS Search aims at helping its users find appropriate scholarly information on the broad topic of social sciences~\cite{DBLP:conf/jcdl/HienertKBZM19}. To this end, it provides different types of information from the social sciences in multiple languages, comprising literature ($114.7$k publications), research data (84k), questions and variables ($13.6$k), as well as instruments and tools ($440$).
A well-configured relevance ranking together with a well-defined structure and faceting mechanism allow to address the users' information needs, however, the most interesting aspect is the inclusion of scientific literature with research data. Typically, those types of information are accessible through different portals only, posing the problem of a lack of links between these two types of information. GESIS Search provides such an integrated access to research data as well as to publications. The information items are connected to each other based on links that are either  manually created or automatically extracted by services that find data references in full texts. Such linking allows researchers to explore the connections between information items interactively.


\subsection{GESIS Search Dataset}

\begin{figure}[t]
\input{sections/gesis-code}
\caption{Examples for publication documents, research dataset documents, and candidate lists for the GESIS Search system.}
\label{listing:gesisdata}
\end{figure}

For LiLAS, we focus on all publications and research data comprised by GESIS Search.
The publications are mostly in English and German, and are annotated with further textual metadata including title, abstract, topic, persons, and others. Metadata on research data comprises (among others) a title, topics, datatype, abstract, collection method, primary investigators, and contributors in English and/or German. 

The data provided to participants comprises the mentioned metadata on social science literature and research data on social science topics comprised in the GESIS Search. In Figure~\ref{listing:gesisdata} we include an excerpt of the different elements included in the data.
For the dataset recommendation task with pre-computed results (see details in Section~\ref{sec:rectask}), in addition, the participants were given the set of research data candidates that are recommended for each publication. This candidate set is computed based on context similarity between publications and research data.
It is created by applying the TF-IDF score to vectorize the combination of title, abstract, and topics for each document type and computing the cosine similarities between cross-data types. It contains a list of research data for each publication with the highest similarities to the publication among other research data in the corpus.


\subsection{Task}\label{sec:rectask}
Research data is of high importance in scientific research, especially when making progress in experimental investigations. However, finding useful research data can be difficult and cumbersome, even if using dataset search engines, such as Google Dataset Search\footnote{\url{https://datasetsearch.research.google.com/}}.
Another approach is scanning scientific publication for utilized or mentioned research data; however, this allows to find explicitly stated research data and not other research data relevant to the subject.
To alleviate the situation, we aim at evolving the recommendation of appropriate research data beyond explicitly mentioned or cited research data. 
To this end, we propose to recommend research data based on publications of the user’s interest between a scientific publication and possible research data candidates. 

The main task is: given a seed-document, participants are asked to calculate the best fitting research data recommendations with regards to the seed-document.
This resembles the use case of providing highly useful recommendations of research data relevant to the publication that the user is currently viewing. For example, the user is interested in the impact of religion on political elections. She finds a publication regarding that topic, which has a set of research data candidates covering the same topic. 

The participants were allowed to submit pre-computed and live runs (see section~\ref{sec:submissiontypes} for more details). For submitting the pre-computed run, the participants also received a first candidate list comprising 1k publication each having a list of recommended research data. The task here was to re-rank this candidate list. 
On the contrary, for submitting the live runs, such a candidate list was not needed, as the recommended candidates needed to be calculated first. To do so, participants are provided metadata on publications as well as on the research data comprised in GESIS Search 
(see Section~\ref{sec:gesissearch} for more details on the provided data).


\section{Evaluation Setup}
\label{sec:setup}

\subsection{STELLA Infrastructure}

\begin{figure}[t]
    \centering
    \includegraphics[scale=0.25]{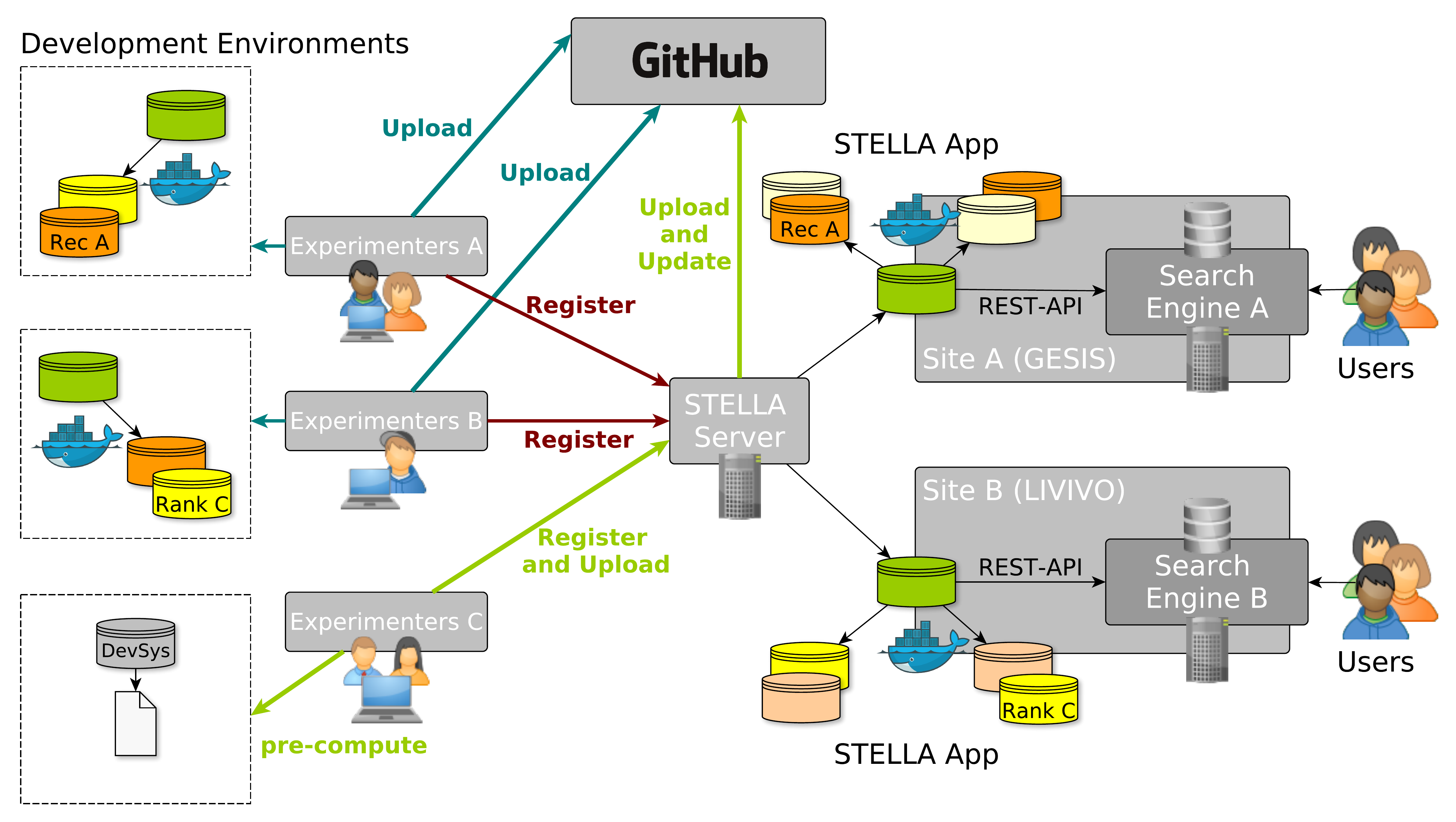}
    \caption{Overview of the STELLA infrastructure}
    \label{fig:infrastructure}
\end{figure}

The technical infrastructure and platform was provided by our evaluation service called STELLA~\cite{breuer2021living} (as illustrated in Figure~\ref{fig:infrastructure}). It complements existing shared task platforms by allowing experimental ranking and recommendation systems to be fully integrated into an evaluation environment, with no interference in the interaction between the users and the system as the whole process is transparent for users. Besides transparency and reproduciblity, one of the STELLA main principles is the integration of experimental systems as micro-services. More specifically, lab participants package their single systems as Docker containers that are bundled in a multi-container application (MCA). Providers of academic research infrastructures deploy the MCA in their back-end and use 
the REST-API either to get ranking and recommendations or to post the corresponding user feedback that is mainly used for our evaluations. Intermediate evaluation results are available through a public dashboard service that is hosted on a central server, also part of the STELLA infrastructure. After authentication, participants can register experimental systems at this central instance and access feedback data that can be used to optimize their systems. In the following, each component of the infrastructure is briefly described to give the reader a better idea on how STELLA serves as a proxy for user-oriented experiments with ranking and recommendation systems.

\subsubsection{Micro-services} 
As pointed out before, we request our lab participants to package their systems with Docker. 
For the sake of compatibility, we provide templates for these micro-services to implement minimal REST-based web services. Participants can adapt their systems to these templates as they see it fits as long as the pre-defined REST endpoints deliver technically correct responses. The templates can be retrieved from GitHub\footnote{\url{https://github.com/stella-project/stella-micro-template}} that is fundamental to our infrastructure. Not only the templates, but also the participant systems should be hosted in a public Git repository in order to be integrated into the MCA. As soon as the developments are done, the participants register their Git(Hub) URL at the central dashboard service of the infrastructure.

\subsubsection{Multi-container Application (MCA)} Once the experimental systems pass technical unit tests and sanity checks for selected queries and target items, they are ready to be deployed and evaluated via user interactions. To reduce the deployments costs for the site providers, the single experimental systems are bundled into an MCA which serves as the entry point to the infrastructure. The MCA handles the query distribution among the experimental systems and also sends user feedback data to the central server at regular intervals. After the REST-API corresponding to the MCA is connected to the search interface, 
the user traffic can be redirected to the MCA which will actually deliver the experimental results. We then interleave results of single experimental systems with those from the baseline system by using a Team-Draft-Interleaving (TDI) approach. This results in two benefits: 1) we prevent users from subpar retrieval results that also might affect the site's reputation, and 2), as shown before, interleaved results can be used to infer statistically significant results with less user data as compared to conventional A/B tests. The site providers rely on their own logging tools. STELLA expects a minimal set of information required when sending feedback; however, sites are free to add any additional JSON-formatted feedback information and interactions to the data payload, for instance logged clicks on site-specific SERP elements. The underlying source code of the MCA is hosted in a public GitHub repository\footnote{\url{https://github.com/stella-project/stella-app}}.

\subsubsection{Central Server} The central server instance of the infrastructure fulfills four functionalities: 1) participants, sites and administrators visit the server to register user accounts and systems; 2) a dashboard service provides visual analytics and first insights about the performance of experimental systems; 3) likewise, feedback data in the form of user interactions is stored in a database that can be downloaded for system optimizations and further evaluations; and 4) the server implements an automated update job of the MCA in order to integrate newly submitted systems if suitable.

Each MCA that is instantiated with legitimate credentials posts the logged user feedback to the central infrastructure server. Even though the infrastructure would allow continuous integration of newly submitted systems, we stuck to the official dates of round 1 and 2 when updating the MCAs at the sites. Due to moderate traffic, we run the central server on a lightweight single core virtual machine with 2GB RAM and 50GB storage capacity\footnote{\url{https://lilas.stella-project.org/}}. More technical details about the implementations can be found in the public GitHub repository\footnote{\url{https://github.com/stella-project/stella-server}}.

\subsection{Submission Types}\label{sec:submissiontypes}

Participants can choose between two different submission types for both tasks (i.e. ad-hoc search and dataset recommendation). Similar to previous living labs, \textbf{Type A} are pre-computed runs that contain rankings and recommendations of the most frequent queries and the most frequently viewed document, respectively for reach task. Alternatively, it is possible to integrate the entire experimental system as a micro-service as part of a \textbf{Type B} submission. Both submission types have their own distinct merits as described below.

\subsubsection{Type A - Pre-computed Runs} Even though the primary goal of the STELLA framework is the integration of entire systems as micro-services, we offer the possibility to participate in the experiments by submitting system outputs, i.e. in the form of pre-computed rankings and recommendations. We do so for two reasons. First, the Type A submissions resemble those of previous living labs and serve as the baseline in order to evaluate the feasibility of our new infrastructure design. Second, we hope to lower technical barriers for some participants that want to submit the system outputs only. To make it easier for participants, we follow the familiar TREC run file syntax.




Depending on the chosen task, for each of the selected top-k queries or target items (identified by \texttt{<qid>}) a ranking or recommendation has to be computed in advance and then uploaded to the dashboard service. The upload process is tightly integrated into the GitHub ecosystem. Once the run file is uploaded, a new repository is automatically created from the previously described micro-template to which the uploaded run is committed. 
This is made possible thanks to GitHub API and access tokens. The run file itself is loaded as a pandas \texttt{DataFrame} into the running micro-service when the \textit{indexing} endpoint is called. Upon request, the queries and target items are translated into the corresponding \texttt{<qid>} to filter the \texttt{DataFrame}. Due to manageable sizes of top-k queries and target items, the entire (compressed) run file can be uploaded to the repository and can be kept in memory after it is indexed as a \texttt{DataFrame}. As a technical safety check, we also integrate a dedicated verification tool\footnote{\url{https://github.com/stella-project/syntax_checker_CLI}} in combination with GitHub Actions to verify that the uploaded files follow the correct syntax.

\subsubsection{Type B - Docker Containers}

Running fully-fledged ranking and recommendation systems as micro-services overcomes the restrictions of responses that are limited to top-k queries and target items. Therefore, we offer the possibility to integrate the entire systems as a Docker container into the STELLA infrastructure as part of Type B submissions. As pointed out earlier, participants fork the template of the micro-services and adapt it to their experimental system. While Docker and the implementation of pre-defined REST endpoints are hard requirements, participants have total freedom w.r.t. the implementation and tools they use within their container, i.e., they do not even have to build up on the Python web application that is provided in the template. Solely, the \textit{index} endpoint and, depending on the chosen task, either the \textit{ranking} or \textit{recommendation} endpoint have to deliver technically correct results. For this purpose, we include unit tests in the template repository that can be run in order to verify that the Docker containers can be properly integrated. If these unit tests pass, the participants register the URL of the corresponding Git repository at the dashboard service. Later on, the system URL is added to the build file of the MCA when an update process is invoked. If the MCA is updated at the sites, newly submitted experimental systems are build from the Dockerfiles in the specified repositories.

\subsection{Baseline Systems}

LIVIVO baseline system for ranking is built on Apache Solr 
and Apache Lucene. 
The index contains about 80 million documents from more than 50 data sources in multiple languages and about 120 searchable fields ranging from basic data such as Title, Abstract, Authors to more specific such as MeSH-Terms, availability or OCR-Data.
For ranking, LIVIVO uses the Lucene default ranker which is a variant of TF-IDF; on top of it, a custom boosting is added. Newer documents as well as search queries occurring in title or author fields are boosted. An exact match of search phrases in title-field  results in a very high boosting. Moreover LIVIVO uses a Lucene-based plugin which executes NLP-tasks like stemming, lemmatization, multilingual search; it also makes use of semantic technologies, mainly based on the Medical Subject Headings (MeSH) vocabulary.

The baseline system for recommendation of research data based on publications in Gesis Search utilizes Pyserini, a Python interface to the IR toolkit built on Lucene designed to support reproducible IR research. The baseline system for recommendation applies the SimpleSearcher of Pyserini that provides the entry point for sparse retrieval  BM25 ranking using bag-of-words representations. 
The Lucene-based index contains abstracts and titles of all research data. The publication identifier (target item of the recommendation) is translated into the publication title, which, in turn, is used to query the index with a BM25 algorithm. Accordingly, the research data recommendations are based on the title and abstracts of the research data and queries made from the publication titles.

\subsection{Evaluation Metrics}
Our logging infrastructure allows us to track search sessions and the corresponding interactions made by users. Each session comprises a specific site user, multiple queries (or target items) as wells as the corresponding results and feedback data in the form of user interactions, primarily logged as clicks with timestamps.

Similar to previous living lab initiatives, we design our user-oriented experiments with interleaved result lists. Given a list with interleaved results and the corresponding clicks of users, we determine \textit{Wins}, \textit{Losses}, \textit{Ties}, and the derived \textit{Outcomes} for relative comparisons of the experimental and baseline systems~\cite{DBLP:conf/clef/SchuthBK15}. Following previous living lab experiments, we implement the interleaving method by the \textit{Team-Draft-Interleaving} algorithm~\cite{DBLP:conf/cikm/RadlinskiKJ08}. More specifically, we refactored exactly the same implementation\footnote{\url{https://bitbucket.org/living-labs/ll-api/src/master/ll/core/interleave.py}} for the highest degree of comparability.

\begin{figure}[t]
    \centering
    \includegraphics[scale=0.2]{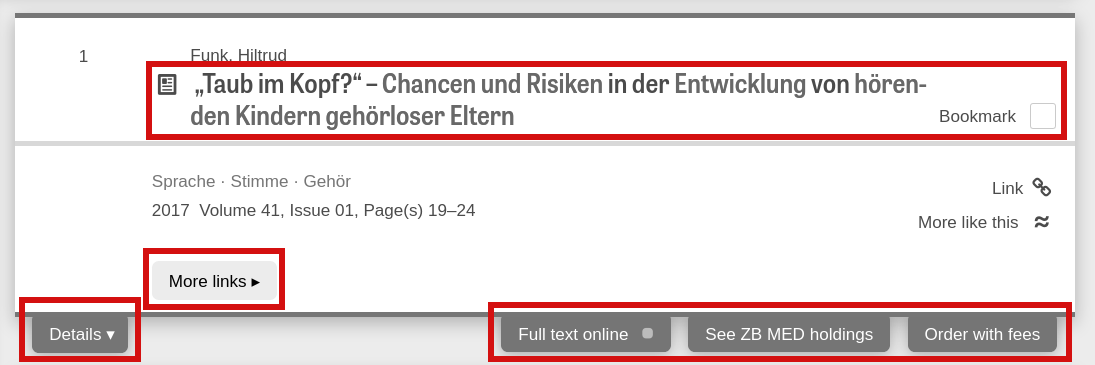}
    \qquad
    \begin{tabular}[b]{lc}
    \toprule
    SERP Element & $w_s$ \\
    \midrule
    Bookmark            &  10 \\
    Order               &  10 \\
    Fulltext            &  8 \\
    In Stock            &  8 \\
    More Links          &  2 \\
    Title               &  1 \\
    Details             &  1 \\
    \bottomrule
    \end{tabular}
    \captionlistentry[table]{Example illustrating the SERP elements for that clicks were logged at LIVIVO and the corresponding weights $w_s$ according to Equation \ref{eq:weighted_sum}.}
    \captionsetup{labelformat=andtable}
    \caption{Example illustrating the SERP elements for that clicks were logged at LIVIVO and the corresponding weights $w_s$ according to Equation \ref{eq:weighted_sum}.}
    \label{tab:livivo_weights}
  \end{figure}

Furthermore we follow Gingstad et al.'s proposal of a weighted score based on click events \cite{DBLP:conf/cikm/GingstadJB20} and define the \textit{Reward} as

\begin{equation}
    \mathrm{\textit{Reward}} = \sum_{s \epsilon S}  w_{s} c_{s}
    \label{eq:weighted_sum}
\end{equation}

where $S$ denotes the set of all elements on a search engine result page (SERP) for which clicks are considered, $w_{s}$ denotes the corresponding weight of the SERP element $s$ that was clicked, and $c_{s}$ denotes the total number of clicks on the SERP element $s$. The \textit{Normalized Reward} is defined as 

\begin{equation}
    \mathrm{\textit{nReward}} = \frac{\mathrm{\textit{Reward}_{exp}}}{\mathrm{\textit{Reward}_{exp}} + \mathrm{\textit{Reward}_{base}}}
\end{equation}

that is the sum of all weighted clicks on experimental results ($\mathrm{\textit{Reward}_{exp}}$) normalized by the total \textit{Reward} given by $\mathrm{\textit{Reward}_{exp}} + \mathrm{\textit{Reward}_{base}}$. Note that, only those clicks from the experimental systems where rankings were interleaved with results of the two compared systems are considered. Figure \ref{tab:livivo_weights} shows the SERP elements that were logged at LIVIVO and the corresponding weights for our evaluations. We do not implement the \textit{Mean Normalized Reward} proposed by Gingstad et al. due to a different evaluation setup. Our lab is organized in rounds during which the systems as well as the underlying document collections are not modified and we already determine the \textit{Normalized Reward} over all aggregated clicks of a specific round.



\subsection{Lab Rounds and Overall Lab Schedule}

The lab was originally split in two separated rounds of 4 weeks each. Due to technical issues for LIVIVO round~1 was four days shorter and round~2 started one week later as planned. To compensate this, we decided to let round~2 last until 24 May 2021, so in total round~2 lasted nearly six instead of four weeks. 
Each participating groups received a set of feedback data after each round; the feedback was also made publicly available on the lab website\footnote{\url{https://th-koeln.sciebo.de/s/OBm0NLEwz1RYl9N}}. Before each round a training phase was offered to allow the participants to build or adapt their systems to the new datasets or click feedback data. 


\section{Participation}
\label{sec:participation}

\subsection{Team lemuren}

Team lemuren participated in both rounds with pre-computed results and dockerized systems for the ad-hoc search task at LIVIVO \cite{tran2021adhoc}. For both rounds, they submitted two different approaches. 


The pre-computed ranking results of \texttt{lemuren elk} are based on built-in functions of Elasticsearch. This system uses a combination between the divergence from randomness model and the Jelinek-Mercer smoothing method for re-ranking candidate documents. The preprocessing pipeline implements stop-word removal, stemming and considers synonyms for medical and  COVID19-related terms. The system was tuned only to the results in English.


\texttt{save fami} is another pre-computed system. It also uses Elasticsearch combined with natural language processing (NLP) modules implemented with the Python package spaCy.
Similar to the second submission for the pre-computed round, this dockerized system is build on top of Elasticsearch and spaCy.
The indexing pipeline follows a multilingual approach supporting English and German languages. 
For both languages the system implements full solutions available in spaCy, either by the models \texttt{en\_core\_sci\_lg} (English biomedical texts) or \texttt{de\_core\_news\_lg} (general German texts).
The system uses the Google Translator API \footnote{\url{https://pypi.org/project/google-trans-new/}} for language detection and automatic translating of incoming queries (from German to English and vice versa).
For indexing and document-retrieval Elasticsearch was used with a custom boosting for MeSH and Chemical-tokens.
\texttt{lemuren elastic only (LEO)} is the second dockerized system by this team which, different from  LEPREP, relies only on Elasticsearchs built-in tools for indexing documents and processing queries.
For indexing documents a custom ingestion pipeline is used to detect the documents language (English or German) and creating the corresponding language fields.
Handling of basic acronyms was modeled by using the built-in word-delimiter function.
Similar to LEPREP-System, LEO uses Google Translator API for automatic query translation.
The system is complemented by a fuzzy match and fuzzy query-expansion to obtain better results for mistyped queries.
Like \texttt{lemuren elk} in round one, LEO also uses DFR and LMJelinekMercer to calculate a score and a similarity distance.

\subsection{Team tekma}

Team tekma contributed experiments to both rounds. In the first round, they submitted the pre-computed results of the system \texttt{tekma\_s} for the ad-hoc search task at LIVIVO \cite{keller2021tekma}. In the second round, they submitted pre-computed recommendations (covering the entire volume of publications) for the corresponding task at GESIS. Both systems are described below.

\texttt{tekma\_s} used Apache Solr to index the document and used pseudo-relevance feedback to extend the queries for the ad-hoc search task. The system only considers documents in English. The system got few impressions and clicks in comparison to the baseline system.
\texttt{ tekma\_n} participated in the second round producing pre-computed recommendations. They used Apache Solr BM25 ranking function and applied query expansion and data enrichment by adding the metadata translations and re-ranking the retrieved result using user feedback and KNN. To generate the primary recommendations for a publication, they used publication fields as a query to search the indexed dataset. 

\subsection{Team GESIS Research}

In addition to the baseline system, team GESIS Research contributed a fully dockerized system in both rounds \cite{tavakolpoursaleh2021pyterrierbased}. 
\texttt{gesis\_rec\_pyterrier} implements a naive content-based recommendation without any advanced knowledge about user preferences and usage metrics. It uses the metadata available in both entity types, i.e., title, abstract, and topics. They employed the classical tfidf-based weighting model from the PyTerrier framework
to obtain first-hand experience with the online evaluation. The indexing and query have been made of the combination of words in title, abstract, and research data topics and publications. They decided to submit the same experimental system for both rounds to gain more user feedback for their unique system. Even though only tfidf-based recommendations are implemented at the current state, it offers a good starting point for further experimentation with PyTerrier and the declarative manner of defining retrieval pipelines. 

\section{Results}
\label{sec:results}


Our experimental evaluations are twofold. First, we evaluate overall statistics of both rounds and sites. Second, we evaluate the performance of all participating systems based on the click data logged during the active periods. As mentioned before, the first round ran during four weeks from March 1st, 2021 to March 28th, 2021 and the second round for five weeks from April 17th, 2021 until May 24th, 2021 at LIVIVO and for six weeks from April 12th, 2021 until May 24th, 2021 at GESIS. 
To foster transparency and reproducibility of the evaluations, we release the corresponding evaluation scripts in an open-source GitHub repository\footnote{\url{https://github.com/stella-project/stella-evaluations}}.

\subsection{Overall evaluations of both rounds and sites}

\begin{table}[t]
\centering
\caption{Number of Sessions, impressions, clicks and click through rate (CTR).}
\begin{tabular}{llcccc}
\toprule
Evaluation round &    Site &  Sessions &  Impressions &  Clicks &    CTR \\
\midrule
         Round 1 &  LIVIVO &      2852 &         4658 &    2452 & 0.5264 \\
         Round 1 &   GESIS &      4568 &         8390 &     152 & 0.0181 \\
         Round 2 &  LIVIVO &     12962 &        25830 &   11562 & 0.4476 \\
         Round 2 &   GESIS &      6576 &        12068 &     250 & 0.0207 \\
\bottomrule
\end{tabular}
\label{tab:overview_round_1_2}
\end{table}

Table \ref{tab:overview_round_1_2} provides an overview of the traffic logged in both rounds. In sum, substantially more sessions, impressions, and clicks were logged in the second round not only due a longer period but also because more systems contributed as Type B submissions. In the first round, systems deployed at LIVIVO were mostly contributed as Type A submissions, meaning their responses were restricted to pre-selected head queries. LIVIVO started the second round with full systems which delivered results for arbitrary queries and thus more session data was logged. GESIS started both rounds with the majority of systems contributed as type B submissions. In comparison to LIVIVO, more sessions and impressions were logged in the first round, but less recommendations were clicked. Similarly, there are less clicks in the second round in comparison to LIVIVO, which is also reflected by the Click-Through Rate (CTR) that is determined by the ratio between Clicks and Impressions. 
As mentioned before, GESIS introduced the recommendations of research datasets as a new service, and, presumably, users were not aware of this new feature.



\begin{figure}[t]
    \centering
    \includegraphics[scale=0.5]{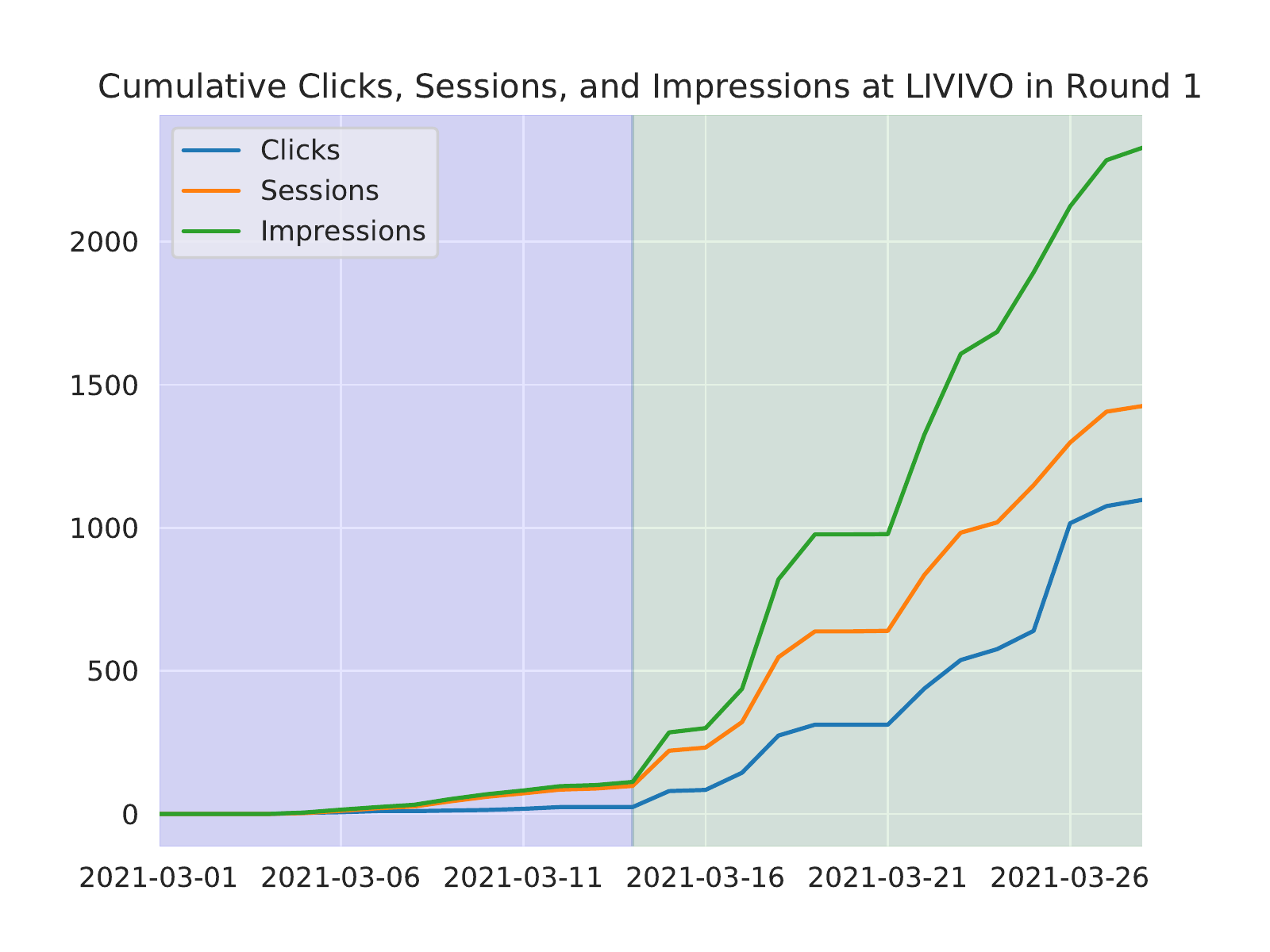}
    \caption{Cumulative sum of logged session data at LIVIVO before (blue) and after (green) the first fully dockerized system went online in the first round.}
    \label{fig:cumsum}
\end{figure}

During the first two weeks of the first round, the amount of logged data at LIVIVO is comparatively low due to systems with pre-computed results for pre-selected head queries. After that, the first type B systems was deployed and increasingly more user traffic could be redirected to our infrastructure. Figure \ref{fig:cumsum} illustrates these effects. The cumulative sums of logged sessions, impressions, and clicks rapidly increased after the first Type B system got online in mid-March.

\begin{figure}[t]
    \centering
    \includegraphics[scale=0.5]{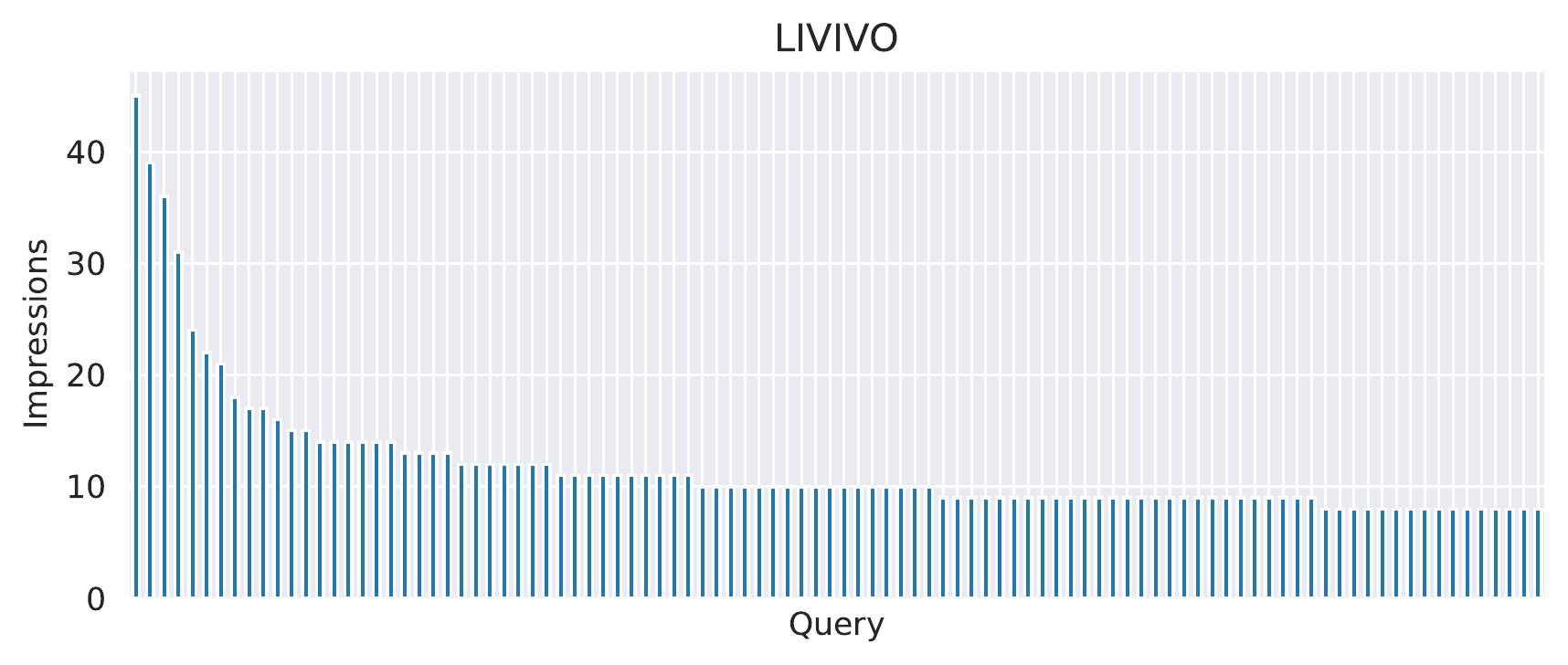}
    \includegraphics[scale=0.5]{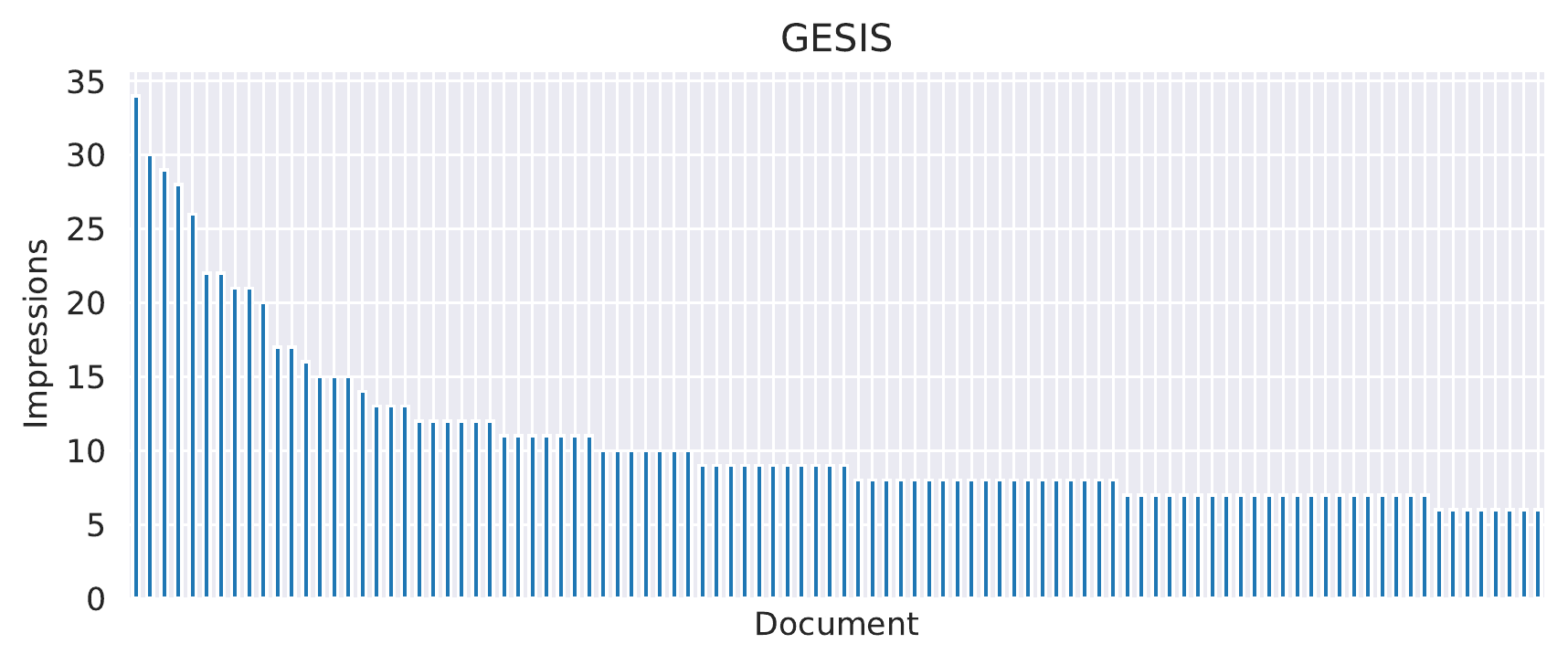}
    \caption{Impressions vs. Query/Document}
    \label{fig:impression_vs_query}
\end{figure}


The logged impressions follow a power-law distribution for both rankings and recommendations as shown in Figure \ref{fig:impression_vs_query}. Most of the impressions can be attributed to a few top-k queries (rankings) or documents (recommendations). The COVID-19 pandemic has a clear influence on the query distributions: the most frequent and the fifth most frequent query are ``covid19'' and ``covid'', respectively. Three of the ten most frequent queries are German queries (``demenz'', ``pflege'', ``schlaganfall''); others are either domain-specific or can be interpreted as English queries. In Table \ref{tab:query_stats_livivo} we report statistics about the queries logged during both rounds at LIVIVO. In both rounds, interaction data was logged for 11,822 unique queries with an average length of 2.9840 terms and each session had 1.9340 queries on average. Nine out of the ten most frequent target items of the recommendations at GESIS are publications with German titles.

\begin{figure}[t]
    \centering
    \includegraphics[scale=0.5]{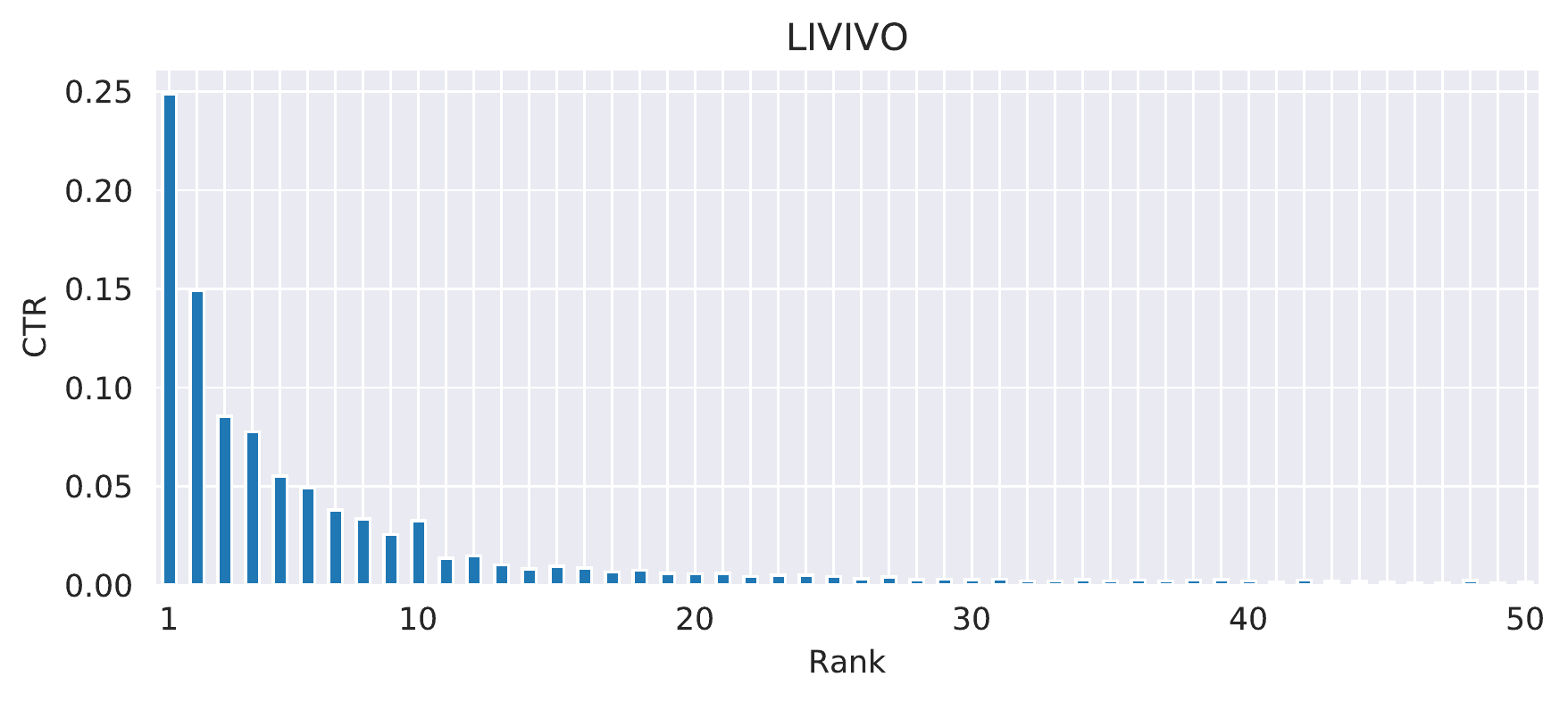}
    \includegraphics[scale=0.5]{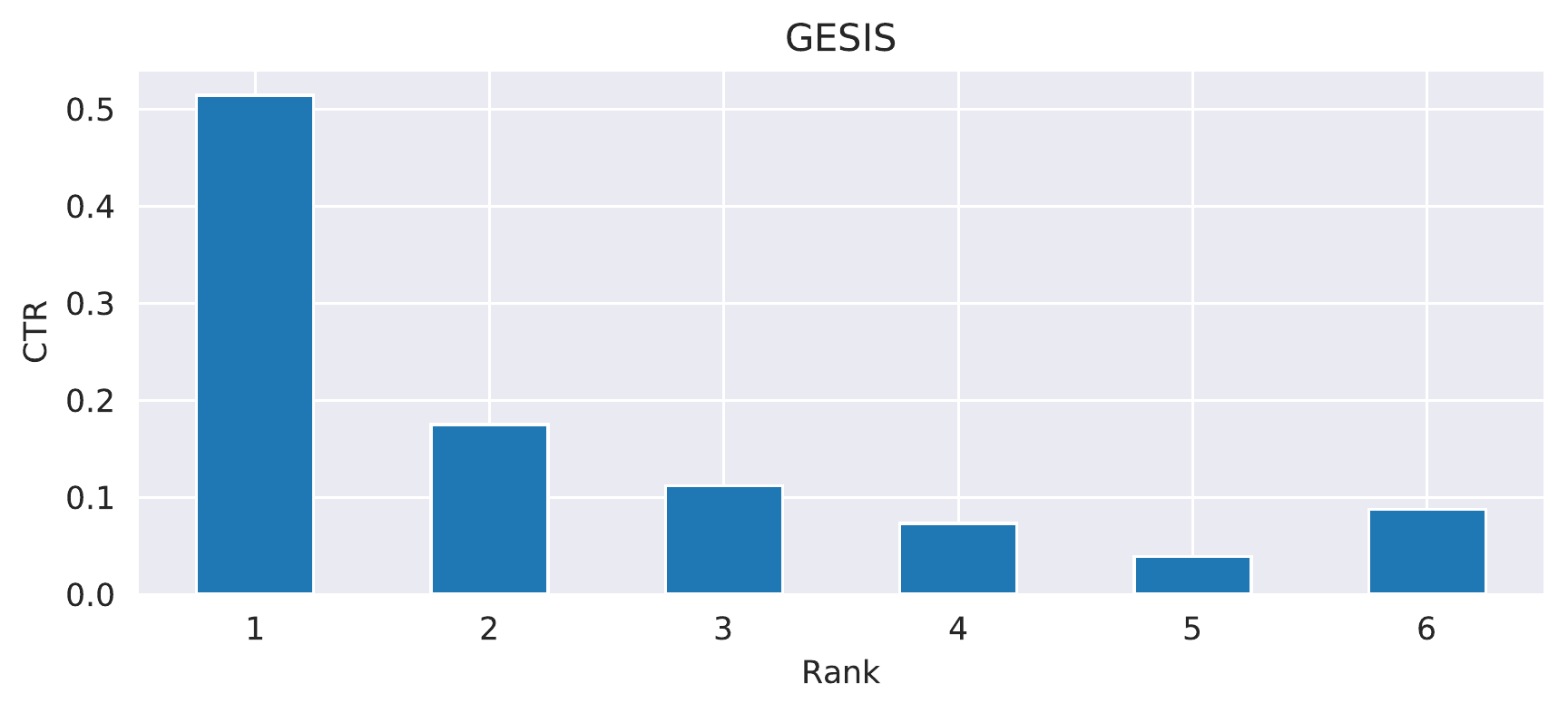}
    \caption{Click-through Rate (CTR) vs. Rank}
    \label{fig:ctr_vs_rank}
\end{figure}

\begin{figure}[t]
    \centering
    \includegraphics[scale=0.5]{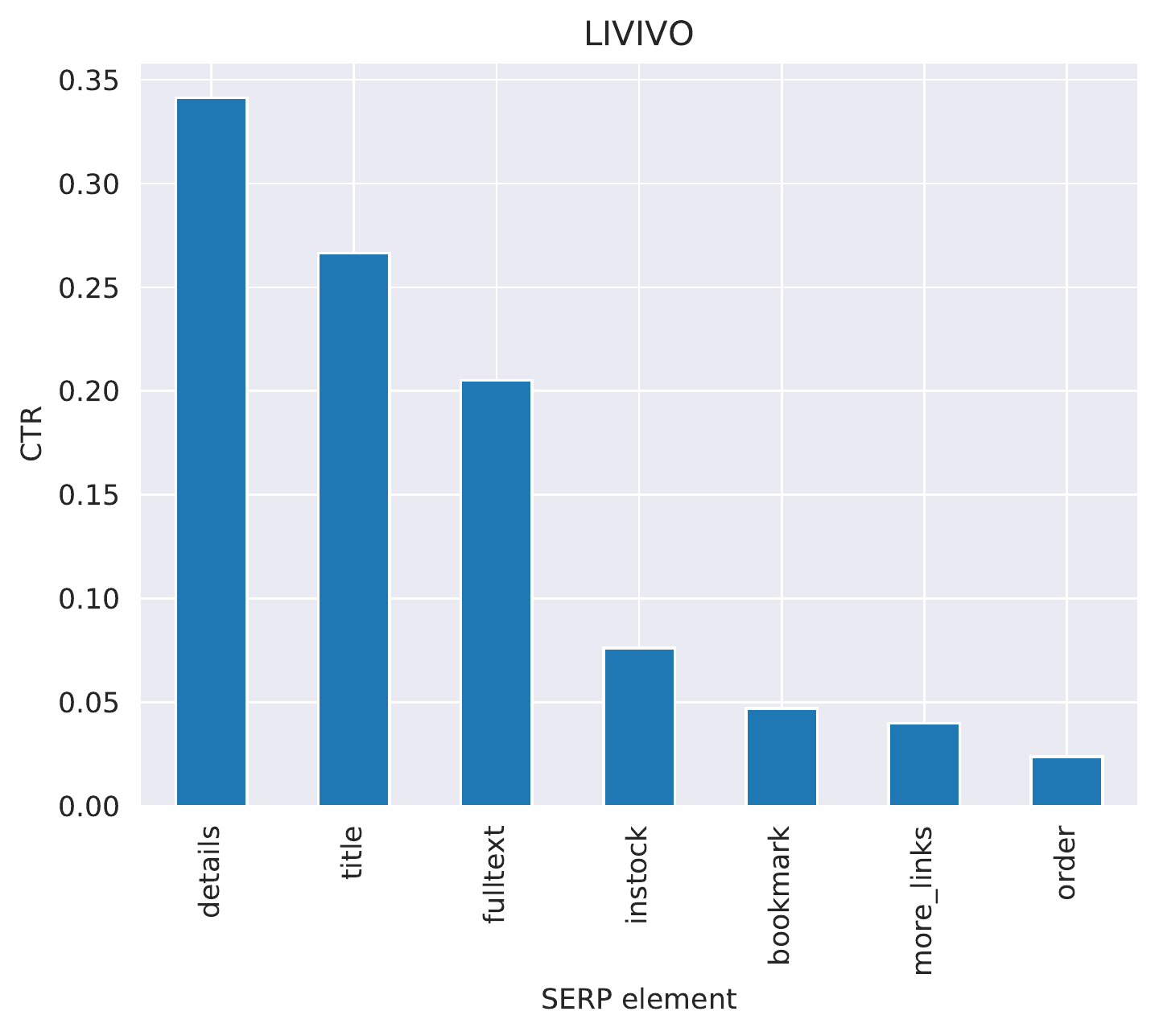}
    \caption{Click distribution on SERP elements at LIVIVO}
    \label{fig:livivo_ctr}
\end{figure}

\begin{table}[t]
\centering
\caption{Statistics of the queries at LIVIVO}
\begin{tabular}{lr}
\toprule
Number of Unique Queries              &  11822 \\
Average Query Length [Terms]          &  2.9840 \\
Average Number of Queries per Session &  1.9340 \\
Average Number of Clicks per Query    &  0.4547 \\
\bottomrule
\end{tabular}
\label{tab:query_stats_livivo}
\end{table}

Another important aspect to be considered as part of the system evaluations is the position bias inherent in the logged data. Click decisions are biased towards the top ranks of the result lists as shown in Figure \ref{fig:ctr_vs_rank}. For both use cases, the rankings and recommendations were displayed to users as vertical lists. Note that, GESIS restricted the recommendations to the first six recommended datasets and no pagination over the following recommended items was possible. LIVIVO shows ten results per page to its users, and as it can be seen from the logged data, users rarely click results beyond the fifth page.

In addition to ``simple'' clicks on ranked items, we logged specific SERP elements that were clicked at LIVIVO. Table \ref{tab:livivo_weights} already provided an overview on which elements were logged and Figure \ref{fig:livivo_ctr} shows the CTR of these elements also follows a power-law distribution. The number of clicks is the highest for the \textit{Details} button and it is followed by the \textit{Title} and \textit{Fulltext} click options. In comparison, the other four logged elements receive substantially less clicks.

\subsection{System evaluations}
An overview of all systems participating in our experiments is provided in Table \ref{tab:system_overview}.
In the first round, three type A systems (\texttt{lemuren\_elk}, \texttt{tekmas}, \texttt{save\_fami}) were submitted and deployed at LIVIVO. They were also deployed in the second round, but did not receive any updates between the two rounds. Since there were no type B submissions in the first round for LIVIVO, we deployed the type B system \texttt{livivo\_rank\_pyserini} after two weeks in mid-March. It provided results for the entire volume of publications and rankings were based on the BM25 method. It was implemented with Pyserini~\cite{DBLP:journals/corr/abs-2102-10073} and the corresponding default settings\footnote{\url{https://github.com/stella-project/livivo_rank_pyserini}}. In contrast to the other systems, it was online for the last two weeks of the first round only. In the second round, it was online in the first days until the other type B systems were ready to be deployed since we wanted to distribute the user traffic among the participants' systems only. In the second round, two type B systems \texttt{lemuren\_elastic\_only} and \texttt{lemuren\_elastic\_preprocessing} were contributed. Both systems build up on Elasticsearch, whereas they differ by the pre-processing as outlined before. At GESIS, \texttt{gesis\_rec\_pyterrier}, submitted as type B system, was online in both rounds. In the first round, the only type A submission was \texttt{gesis\_rec\_precom} that was substituted in the second round by \texttt{tekma\_n}. Both baseline systems at LIVIVO (\texttt{livivo\_base}) and GESIS (\texttt{gesis\_rec\_pyserini}) were integrated as type B systems, remained unmodified, and could deliver results for every request.

\begin{table}[t]
\centering
\caption{System overview}
\begin{tabular}{lccccc}
\toprule
System name & Task & Type & Experimental & Round 1 & Round 2 \\
\midrule
lemuren\_elk & 1 & A & \CIRCLE & \CIRCLE & \CIRCLE \\
tekmas & 1 & A & \CIRCLE & \CIRCLE & \CIRCLE \\
save\_fami & 1 & A & \CIRCLE & \CIRCLE & \CIRCLE \\
livivo\_rank\_pyserini & 1 & B & \CIRCLE & \RIGHTcircle & \RIGHTcircle \\
lemuren\_elastic\_only & 1 & B & \CIRCLE & \Circle & \CIRCLE \\
lemuren\_elastic\_preprocessing & 1 & B & \CIRCLE & \Circle & \CIRCLE \\
livivo\_base & 1 & B & \Circle & \CIRCLE & \CIRCLE \\
tekma\_n & 2 & A & \CIRCLE & \Circle & \CIRCLE \\
gesis\_rec\_precom & 2 & A & \CIRCLE & \CIRCLE & \Circle \\
gesis\_rec\_pyterrier & 2 & B & \CIRCLE &  \CIRCLE & \CIRCLE \\
gesis\_rec\_pyserini & 2 & B & \Circle & \CIRCLE  & \CIRCLE \\
\bottomrule
\end{tabular}
\label{tab:system_overview}
\end{table}


\begin{table}[t]
\centering
\caption{Outcomes of Round 1. Dagger symbols ($\dagger$) indicate baseline systems. Significant differences are denoted by an asterisk symbol ($\ast$).}
\label{tab:round_1}
\begin{tabularx}{\textwidth}{p{4.25cm}XXXXXXXX}
\toprule
System &  \rot{90}{Win} &  \rot{90}{Loss} &  \rot{90}{Tie} & \rot{90}{Outcome} &  \rot{90}{Sessions} &  \rot{90}{Impressions} &  \rot{90}{Clicks} &     \rot{90}{CTR} \\
\midrule
gesis\_rec\_pyserini$\dagger$            &   36 &    36 &    1 &    0.50 &      2284 &         4195 &      37 &  0.0088 \\
gesis\_rec\_pyterrier           &   26 &    28 &    1 &    0.48 &      1968 &         3675 &      28 &  0.0076 \\
gesis\_rec\_precom              &   10 &     8 &    0 &    0.56 &       316 &          520 &      11 &  0.0212 \\
\midrule
livivo\_base$\dagger$                    &  332 &   234 &   67 &    0.59 &      1426 &         2329 &     677 &  0.2907 \\
livivo\_rank\_pyserini          &  215 &   302 &   64 &    $0.42^{\ast}$ &      1260 &         2135 &     517 &  0.2422 \\
lemuren\_elk                   &    4 &     8 &    1 &    0.33 &        45 &           55 &      10 &  0.1818 \\
tekmas                        &    6 &    10 &    1 &    0.38 &        64 &           77 &       8 &  0.1039 \\
save\_fami                     &    9 &    12 &    1 &    0.43 &        57 &           62 &      14 &  0.2258 \\
\bottomrule
\end{tabularx}
\end{table}

Table \ref{tab:round_1} compares the experimental systems' outcomes and the corresponding logged interactions and session data during the first round. Regarding the \textit{Outcome} measure, none of the experimental systems was able to outperform the baseline systems. Note that the reported \textit{Outcomes} of the baseline systems result from comparisons against all experimental systems. The systems with pre-computed rankings (type A submissions) received a total number of 32 clicks over a period of four weeks at LIVIVO. Since interaction data was sparse in the first round, we only received enough data for \texttt{livivo\_rank\_pyserini} to conduct significance tests. The reported p-value results from a Wilcoxon signed-rank test and shows a significant difference between the experimental and baseline system.

Table \ref{tab:round_2} shows the results of the second round. \texttt{tekma\_n} was contributed as type A submission, but results were pre-computed for the entire volume of publications at GESIS. It replaced \texttt{gesis\_rec\_precom} and achieved a higher CTR compared to the other recommender systems. Likewise, it achieves an \textit{Outcome} of $0.62$, which might be an indicator that it outperforms the baseline recommendations given by \texttt{gesis\_rec\_pyserini}. Unfortunately, we are not able to conduct any meaningful significance tests due to the sparsity of click data. At LIVIVO, the systems with pre-computed rankings (type A submissions) received a comparable amount of clicks similar to the first round. In sum, all three systems received a total number of 35 clicks over a period of five weeks. Even though, click data is sparse and interpretations have to be made carefully, the relative ranking order of these three systems is preserved in the second round (e.g. in terms of the \textit{Outcome}, total number of clicks, or CTR). 

In the second round, no experimental system could outperform the baseline system at LIVIVO. Both experimental type B systems \texttt{lemuren\_elastic\_only} and \texttt{lemuren\_elastic\_preprocessing} achieve significantly lower \textit{Outcome} scores as the baseline. However, the second system has substantially lower \textit{Outcome} and CTR scores. Both systems share a fair amount of the same methodological approach and only differ by the processing of the input text. In this case, the system performance does not seem to benefit from this specific pre-processing step, when interpreting clicks as positive relevance signals. The third type B system at LIVIVO \texttt{livivo\_rank\_pyserini} did not participate the entire second round, since we took it offline as soon as the other type B systems were available. 
Despite having participated in comparatively less experiments than in the first round (1260 sessions vs. 243 sessions), the system achieves in both rounds comparable results in terms of \textit{Outcome} and CTR scores. This circumstance raises the question for how long systems have to be online to deliver reliable performance estimates. 

Previous studies showed that a system is more likely to win if its documents are ranked at higher positions~\cite{DBLP:journals/jdiq/JagermanBR18DBLP}. As part of our experimental evaluations, we can confirm this circumstance. We also determined the Spearman correlation between an interleaving outcome (1: win, -1: loss, 0: tie) and the highest ranked position of a document contributed by an experimental system. At both sites, we see a weak but significant correlation (LIVIVO: $\rho=-0.0883$, $p=1.3535e-09$; GESIS: $\rho=-0.3480$, $p=4.7422e-07$).

One shortcoming of the previous measures derived from interleaving experiments is the simplified interpretation of click interactions. As outlined in Section \ref{sec:setup}, by weighting clicks differently, it is possible to account for the meaning of the corresponding SERP elements. Table \ref{tab:round02_reward} shows the total number of clicks on SERP elements for each systems and the \textit{Normalized Reward} (nReward) resulting from the weighting scheme given in Figure \ref{tab:livivo_weights}. We compare the total number of clicks of those (interleaving) experiments in which the experimental and baseline systems delivered results. As it can be seen, comparing systems by clicks on different SERP elements, provides a more diverse analysis. For instance, some of the systems achieve higher numbers of clicks (and CTRs) for some SERP elements in direct comparison to the baseline systems. \texttt{livivo\_rank\_pyserini}, \texttt{lemuren\_elastic\_only} got more clicks on the \textit{Bookmark} element than the baseline system, while all systems achieve lower numbers of total clicks. 

None of the systems could outperform the baseline system in terms of the nReward measure, but in comparison to the \textit{Outcome} scores, there is a more balanced ratio between the nReward scores that also accounts for the meaning of specific clicks. Likewise, it accounts for clicks even if the experimental system did not ``win'' in the interleaving experiment. In Table \ref{tab:round02_reward} we compare the total number of clicks over multiple sessions. While the Win, Loss, Tie, and Outcome only measure if there have been more clicks in a single experiment, the nReward also considers those clicks that were made in experiments in which the experimental system did not necessarily win.

\begin{table}[t]
\caption{Outcomes of Round 2. Dagger symbols ($\dagger$) indicate baseline systems. Significant differences are denoted by an asterisk symbol ($\ast$).}
\centering
\begin{tabularx}{\textwidth}{p{4.25cm}XXXXXXXX}
\toprule
System &  \rot{90}{Win} &  \rot{90}{Loss} &  \rot{90}{Tie} & \rot{90}{Outcome} &  \rot{90}{Sessions} &  \rot{90}{Impressions} &  \rot{90}{Clicks} &     \rot{90}{CTR} \\
\midrule
gesis\_rec\_pyserini$\dagger$             &    51 &    68 &    2 &    0.43 &      3288 &         6034 &      53 &  0.0088 \\
gesis\_rec\_pyterrier           &    26 &    25 &    1 &    0.51 &      1529 &         2937 &      27 &  0.0092 \\
tekma\_n                       &    42 &    26 &    1 &    0.62 &      1759 &         3097 &      45 &  0.0145 \\
\midrule
livivo\_base$\dagger$                    &  2447 &  1063 &  372 &    0.70 &      6481 &        12915 &    3791 &  0.2935 \\
livivo\_rank\_pyserini          &    48 &    71 &   15 &    0.40 &       243 &          434 &     112 &  0.2581 \\
lemuren\_elastic\_only          &   707 &  1042 &  218 &    $0.40^{\ast}$ &      3131 &         6274 &    1273 &  0.2029 \\
lemuren\_elastic\_preprocessing &   291 &  1308 &  135 &    $0.18^{\ast}$ &      2948 &         6026 &     570 &  0.0946 \\
lemuren\_elk                   &     6 &    13 &    0 &    0.32 &        61 &           69 &      10 &  0.1449 \\
tekma\_s                        &     4 &     7 &    1 &    0.36 &        36 &           42 &       5 &  0.1190 \\
save\_fami                     &     7 &     6 &    3 &    0.54 &        62 &           70 &      20 &  0.2857  \\
\bottomrule
\end{tabularx}
\label{tab:round_2}
\end{table}



\begin{table}[t]
\caption{Experimental systems of round 2 and the corresponding number of clicks on SERP elements, total number of clicks, and the \textit{Reward} score.}
\centering
\begin{tabularx}{\textwidth}{lXXXXXXXXX}
\toprule
{} &  \rot{90}{Bookmark} &  \rot{90}{Details} &  \rot{90}{Fulltext} &  \rot{90}{In Stock} &  \rot{90}{More Links} &  \rot{90}{Order} &  \rot{90}{Title} &  \rot{90}{Total Clicks} &    \rot{90}{nReward} \\
\midrule
livivo\_rank\_pyserini               &       182 &      341 &       176 &       55 &          62 &     28 &    263 &    1107 &   0.4367 \\
livivo\_base          &       180 &      443 &       228 &      154 &          57 &     29 &    329 &    1420 &   0.5633 \\
\midrule
lemuren\_elastic\_only               &        63 &      832 &       481 &      107 &         105 &     54 &    638 &    2280 &   0.4045 \\
livivo\_base          &        56 &     1066 &       646 &      295 &         129 &     85 &    858 &    3135 &   0.5955 \\
\midrule
lemuren\_elastic\_preprocessing      &        23 &      355 &       257 &       23 &          28 &     21 &    285 &     992 &   0.2143 \\
livivo\_base &        69 &     1190 &       762 &      301 &         119 &     82 &    934 &    3457 &   0.7857 \\
\midrule
lemuren\_elk                        &         1 &       13 &        16 &        0 &           2 &      0 &     10 &      42 &   0.4242 \\
livivo\_base                   &         1 &       24 &         7 &       14 &           1 &      0 &     20 &      67 &   0.5758 \\
\midrule
tekmas                             &         2 &       11 &         2 &        2 &           1 &      0 &      6 &      24 &   0.3430 \\
livivo\_base                       &         0 &       13 &         6 &        7 &           0 &      1 &      9 &      36 &   0.6570 \\
\midrule
save\_fami                          &        11 &       21 &         9 &        3 &           1 &      1 &     16 &      62 &   0.5496 \\
livivo\_base                     &         8 &       13 &         7 &        5 &           2 &      1 &      6 &      42 &   0.4504 \\
\midrule
All experimental systems                         &       282 &     1573 &       941 &      190 &         199 &    104 &   1218 &    4507 &   0.3485 \\
livivo\_base                        &       314 &     2749 &      1656 &      776 &         308 &    198 &   2156 &    8157 &   0.6515 \\
\bottomrule
\end{tabularx}
\label{tab:round02_reward}
\end{table}

\section{Conclusions}
\label{sec:conclusion}




The Living Labs for Academic Search (LiLAS) lab re-introduced the living lab paradigm with a focus on tasks in the domain of academic search. The lab offered the possibility to participate in two different tasks, which were either dedicated to ad-hoc search in the Life Sciences or research data recommendations in the Social Sciences. Participants were provided with datasets and access to the underlying search portals for experimentation. For both tasks, participants could contribute their experimental systems either by pre-computed outputs for selected queries (or target items) or as fully-fledged dockerized systems. In total, we evaluated nine experimental systems out of which seven were contributed by three participating groups. In sum, two groups contributed experiments that cover pre-computed rankings and fully dockerized systems at LIVIVO and pre-computed recommendations at GESIS. The GESIS research team contributed another completely dockerized recommendation system. Our experimental setup is based on interleaving experiments that combine experimental results with those from the corresponding baseline systems at LIVIVO and GESIS. In accordance with the living lab paradigm, our evaluations are based on user interactions, i.e. in the form of click feedback.

A key component of the underlying infrastructure is the integration of experimental ranking and recommendation systems as micro-services that are implemented with the help of Docker. The LiLAS lab was the first test-bed to use this evaluation service and it exemplified some of the benefits resulting from the new infrastructure design. First of all, completely dockerized systems can overcome the restrictions of results limited to filtered lists of top-k queries or target items. Significantly more data and click interactions can be logged if the experimental systems can deliver results on-the-fly for arbitrary requests of rankings and recommendations. As a consequence, this allows much more data aggregation in a shorter period of time and provides a solid basis for statistical significance tests.

Furthermore, the deployment effort for site providers and organizers is considerably reduced. Once the systems are properly described with the corresponding Dockerfile, they can be rebuild on purpose, exactly as the participants and developers intended them to be. Likewise, the entire infrastructure service can be migrated with minimal costs due to Docker. However, we hypothesize that one reason for the low participation might be the technical overhead for those who were not already familiar with Docker. On the other hand, the development efforts pay off. If the systems are properly adapted to the required interface and the source code is available in a public repository, the (IR) research community can rely on these artifacts that make the experiments transparent and reproducible.

Thus, we address the reproducibility of these living lab experiments mostly from a technological point of view, in the sense that we can repeat the experiments in the future with reduced efforts, since the participating systems are openly available and should be reconstructible with the help of the corresponding Dockerfiles. Future work should investigate how feasible it is to rely on the Dockerfiles for the long-term preservation. Since experimental systems are rebuilt each time with the help of the Dockerfile, updates of the underlying dependencies might be a threat to the reproducibility. An intuitive solution would be the integration of pre-built Docker images that may allow a longer reproducibility. Apart from the underlying technological aspects, the reproducibility of the actual experimental results has to be investigated. Our experimental setup would allow to answer questions with regard to the reproducibility of the experimental results over time and also across different domains (e.g. Life vs. Social Sciences).

Most of the evaluation measures are made for interleaving experiments that also depend on the results of the baseline system and not solely on those of an experimental system. We have not investigated yet, if the experimental results follow a transitive relation: if the experimental system A outperforms the baseline system B, denoted as $A \succ B$, and the baseline system B outperforms another experimental system C ($B \succ C$), can we conclude that system A would also outperform system C ($A \succ C$)? As the evaluations showed, click results are heavily biased towards the first ranks and likewise they are context-dependent, i.e. they depend on the entire result list and single click decisions have to be interpreted in relation to neighboring and previously seen results and further evaluations in these directions would require counterfactual reasoning. Nonetheless, in the second round it was illustrated how our infrastructure service can be used for incremental developments and component-wise analysis of experimental systems. The two experimental systems \texttt{lemuren\_elastic\_only} and \texttt{lemuren\_elastic\_preprocessing} follow a similar approach and only differ by the pre-processing component that has been shown not to be of any benefit.

In addition to established outcome measures of interleaving experiments (Win, Loss, Tie, Outcome), we also account for the meaning of clicks on different SERP elements. In this context, we implement the Reward measure that is the weighted sum of clicks on different elements corresponding to a specific result. Even though most of the experimental systems could not outperform the baseline systems in terms of the overall scores, we see some clear differences between the system performance, which allow us to assess a system's merits more thoroughly, when the evaluations are based on different SERP elements.

Overall, we consider our lab as a successful advancement to previous living lab experiments. We were able to exemplify the benefits of fully dockerized systems delivering results for arbitrary results on-the-fly. Furthermore, we could confirm several previous findings, for instance the power laws underlying the click distributions. Additionally, we were able to conduct more diverse comparison by differentiating between clicks on different SERP elements and accounting for their meaning. Unfortunately, we could not attract many participants, leaving some aspects not tested, e.g. how many systems/experiments can be run simultaneously considering the limitations of the infrastructure design, hardware requirements, server load and user traffic. Likewise, no experimental ranking system could outperform the baseline system. In the future, it might be helpful to provide participants with open and more transparent baseline systems they can build upon. Some of the pre-computed experimental ranking and recommendations seem to deliver promising results; however, the evaluations need to be interpreted with care due to the sparsity of the available click data. As a way out, we favor continuous evaluations freed from the time limits of rounds, in order to re-frame the introduced living lab service as an ongoing evaluation challenge. The corresponding source code can be retrieved from a public GitHub project\footnote{\url{https://github.com/stella-project}} and we plan to release the aggregated session data as a curated research dataset.

\subsubsection*{Acknowledgments} This paper is supported by DFG (project no. 407518790).

\bibliographystyle{splncs04}
\bibliography{bibliography} 

\end{document}